\documentclass[useAMS,usenatbib,usegraphicx]{mn2e}

\newcommand{\taut}{\tau_{\rm T}}

\title[Reflection in BHB systems]
  {X-ray reflection in accreting stellar-mass black hole systems} 
\author[R.R. Ross \& A.C. Fabian]
  {R.~R.~Ross$^1$\thanks{rross@holycross.edu} and A.~C.~Fabian$^2$\\
$^1$Physics Department, College of the Holy Cross, Worcester, MA 01610, USA \\
   $^2$Institute of Astronomy, Madingley Road, Cambridge CB3 0HA}

\pagerange{\pageref{firstpage}--\pageref{lastpage}}
\pubyear{2005}

\usepackage{times}

\begin{document}

\label{firstpage}

\maketitle

\begin{abstract}
The X-ray spectra of accreting stellar-mass black hole systems exhibit spectral 
features due to reflection, especially broad iron K$\alpha$ emission lines. 
We investigate the reflection by the accretion disc that can be expected in the
high/soft state of such a system. First, we perform a self-consistent calculation 
of the reflection that results from illumination of a hot, inner portion of the 
disc with its atmosphere in hydrostatic equilibrium. Then we present reflection 
spectra for a range of illumination strengths and disc temperatures under the
assumption of a constant-density atmosphere. Reflection by a hot accretion
disc differs in important ways from that of a much cooler disc, such as that
expected in an active galactic nucleus.
\end{abstract}


\section{Introduction}

Studies of the spectra of accreting black holes are important for 
understanding the accretion flow and geometry, and hence the effects of 
strong gravity and black-hole spin. In a black-hole binary (BHB) system,
radiatively efficient accretion through a disc generates a quasi-thermal 
spectrum which is well seen in the high/soft states (see Remillard \& 
McClintock 2006). Modeling of such spectra, with detailed attention 
paid to spectral hardening by radiation-transfer effects, has been 
undertaken by Shimura \& Takahara 1995, Merloni, Fabian \& Ross 2000,
and recently by Davis and collaborators (Davis et al.\ 2005; Davis \&
Hubeny 2006).
At the same time, coronal emission in such sources generates a power-law 
spectrum by inverse Compton scattering of the thermal disc photons. This 
hard emission can irradiate the disc to produce a reflection spectrum 
(Guilbert \& Rees 1988; Lightman \& White 1988; George \& Fabian 1991;
Ross \& Fabian 1993) which may contain sharp spectral features that 
can reveal the velocity of the matter and the depth of the gravitational 
well (Fabian et al.\ 2000; Reynolds \& Nowak 2003; Miller 2007).

Here we investigate reflection spectra in BHBs.  Broad iron emission
lines have been found in a wide range of BHBs (Ba\l uci\'{n}ska-Church
\& Church 2000; Martocchia et al.\ 2002; Miller et al.\ 2002a,b,c;
Miller et al.\ 2004; Miniutti, Fabian \& Miller 2004; Rossi et al.\
2005; Miller et al 2006) indicating the presence of reflection
components in those objects, particularly in the very high and the
low/hard states.  Our previous studies of active galactic nuclei (Ross
\& Fabian 1993; Ballantyne, Ross \& Fabian 2001; Ross \& Fabian 2005)
concentrated on the irradiation of slabs of gas which would otherwise
be cold. For a BHB, however, the gas in the accretion disc is expected
to be warm ($T\sim 10^6\,{\rm K}$). Low-$Z$ elements are stripped of
electrons even before photoionization by illuminating radiation is
considered, and thermal emission by the gas can dominate portions of
the X-ray spectrum.

We begin by considering reflection by an accretion-disc surface layer which has
a density structure consistent with the condition of hydrostatic equilibrium. 
We then show a range of spectra produced by constant-density slabs heated from 
below by blackbody radiation and illuminated from above by hard power-law radiation.

\section{Illuminated disc atmosphere}

\subsection{Method}

We begin with a model for illumination of gas in hydrostatic equilibrium 
atop a spatially thin accretion disc. We take the mass of the central black 
hole to be $M=10M_{\odot}$ and the accretion rate to be a fraction 
$\varepsilon=\dot M/\dot M_{\rm Edd}$ of the value corresponding to the 
Eddington limit. If all gravitational energy is dissipated within the
disc, the total flux emerging from the disc at radius $R$ 
(Shakura \& Sunyaev 1973) is expected to be
\begin{equation}
F_{\rm disc}=\frac{3GM\dot M}{8\pi R^3}\left(1-\sqrt{\frac{R_{\rm min}}{R}}\right).
\end{equation}
For a nonrotating black hole, $R_{\rm min}=6R_{\rm g}$, where
$R_{\rm g}=GM/c^2$ is the gravitational radius.

We model the radiative transfer and hydrostatic equilibrium throughout a 
surface layer of total Thomson depth $\taut=10$.  
The base of this surface layer is fixed at height $H$ above the midplane of
the disc, where $H$ is the estimate for the half-thickness of the accretion 
disc given by Merloni et al.\ (2000). The total flux $F_{\rm disc}$ enters  
the surface layer from below in the form of a blackbody spectrum.

The method for treating the radiative transfer and finding the 
self-consistent temperature and ionization state of the gas has been 
described by Ross \& Fabian (2005) and references therein. The 
steady-state radiation field is assumed to satisfy the 
Fokker-Planck/diffusion equation,
\begin{equation}
\left(\frac{\partial n}{\partial t}\right)_{\rm FP}
+\frac{\partial}{\partial z}\left(\frac{c}{3\kappa}
\frac{\partial n}{\partial z}\right)
+\frac{j_Eh^3c^3}{8\pi E^3} - c\kappa_{\rm A}n \equiv 0\,,
\end{equation}
where $n$ is the photon occupation number, $E$ is the photon energy,
$z$ is the vertical height above the midplane of the accretion
disc, $j_E$ is the spectral emissivity, 
$\kappa=\kappa_{\rm A}+\kappa_{\rm KN}$ is the total opacity (per volume), 
$\kappa_{\rm A}$ is the absorption opacity, and $\kappa_{\rm KN}$ is 
the Klein-Nishina opacity for Compton scattering.  The leading term,
$(\partial n/\partial t)_{\rm FP}$, is given by the Fokker-Planck equation
of Cooper (1971) and accurately treats the effects of Compton scattering for 
$E \la 1\,{\rm  MeV}$ and $kT \la 100\,{\rm keV}$. As the radiation field is 
relaxed to a steady state, the local temperature 
and fractional ionization of the gas are found by solving the equations of 
thermal and ionization equilibrium. In addition to fully-ionized species, 
the following ions are included in the calculations:
C\,{\sc iii--vi}, N\,{\sc iii--vii}, O\,{\sc iii--viii}, Ne\,{\sc iii--x}, 
Mg\,{\sc iii--xii}, Si\,{\sc iv--xiv}, S\,{\sc iv--xvi}, and Fe\,{\sc vi--xxvi}.
Elemental abundances are taken from Morrison \& McCammon (1983).
Ross \& Fabian (2005) have summarized the atomic data employed.
In particular, it should be noted that three-body recombination, which
can be important for elements lighter than iron at the high densities
in our models, is included in the recombination rate tables of Summers (1974) 
that are employed. Three-body recombination of iron is neglected, since it is 
only important at higher densities ($n_{\rm e}\ga 10^{22}\,{\rm cm}^{-3}$)
than occur in our models (Jacobs et al.\ 1977).

The density structure of the gas within the surface layer is found from
the condition for hydrostatic equilibrium:
\begin{equation}
-\frac{dP_{\rm gas}}{dz} -\frac{1}{3}\frac{du}{dz} 
-\frac{GM\rho}{R^3}z=0,
\end{equation}
where $P_{\rm gas}$ is the gas pressure, $\rho$ is the gas density, and
$u$ is the total radiation energy density. Details of the method have been 
described by Ballantyne et al.\ (2001).

\subsection{Results}
   
We consider an illuminated accretion-disc surface located at $R=12R_{\rm g}$ 
when the accretion rate corresponds to $\varepsilon=0.1$ times the Eddington 
limit. The blackbody radiation emerging from the disc then corresponds to a
temperature such that $kT_{\rm BB}\approx 0.35\,{\rm keV}$. The outer 
surface is illuminated by a cutoff power-law spectrum with photon index 
$\Gamma = 2$ and $e$-folding energy $E_{\rm cut} = 300\,{\rm keV}$ for the 
high-energy exponential cutoff. For self-consistent results, photon energies 
ranging from 1~eV to 1~MeV are treated in the radiative transfer calculation. 
However, the illuminating spectrum, which is probably produced by Compton 
upscattering of disc radiation by a hot accretion-disc corona (e.g., Shapiro,
Lightman \& Eardley 1976; Haardt \& Maraschi 1993; Dove, Wilms \& Begelman 1997), 
cannot be expected to extend to extremely low energies. Therefore, the 
illuminating spectrum is given an abrupt low-energy cutoff at $E = 0.1\,{\rm keV}$. 
The illumination has a total flux $F_0 = 0.1F_{\rm disc}$, so that the thermal 
radiation dominates. (In the  2--20~keV range, the illuminating flux is 0.17 times 
the disc flux.)
 
\begin{figure}
\includegraphics[scale=0.34,angle=270]{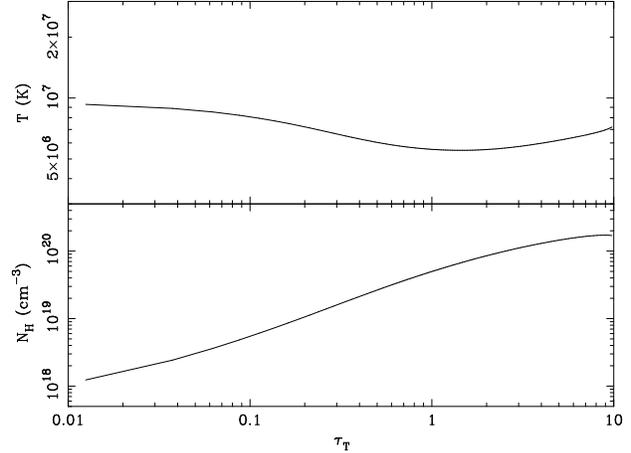}
\caption{Temperature (top) and hydrogen number density (bottom) as functions
of Thomson depth when the surface of a particular black-hole accretion disc at 
$R=12R_{\rm g}$ is illuminated by a cutoff power-law spectrum with total flux
$F_0 = 0.1F_{\rm disc}$.} 
\end{figure}
The calculated temperature and density structures of the surface layer are shown 
in Figure~1. The thermal radiation from the disc below keeps the temperature
high throughout the surface layer, and the external radiation causes the 
temperature to rise with decreasing Thomson depth for $\taut\la 1$. Elements
lighter than iron are fully ionized throughout the surface layer. The disc
radiation is too soft to ionize iron K-shell electrons, however, and Fe\,{\sc xxv}
dominates for $\taut\ga 1/3$. The gas density decreases steadily as $\taut$
decreases. There is no abrupt change in temperature and density like that found
when a cooler AGN accretion disc is illuminated (Nayakshin, Kazanas \& Kallman 2000;
Ballantyne et al.\ 2001; R\'{o}\.{z}a\'{n}ska et al.\ 2002).

\begin{figure}
\includegraphics[scale=0.34,angle=270]{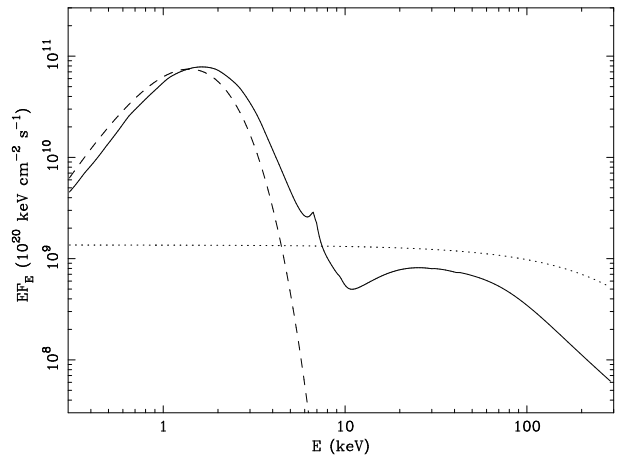}
\caption{Spectrum (solid curve) emerging from the illuminated accretion-disk surface
with the atmosphere in hydrostatic equilibrium. The blackbody spectrum entering the 
surface layer from below and the cutoff power-law spectrum illuminating the outer 
surface are represented by the dashed and dotted curves, respectively.} 
\end{figure}
The spectrum emerging from the surface of the disc is shown in Figure~2. The hotter
gas near the illuminated surface causes the 2--10~keV ``thermal'' spectrum to be 
enhanced both by Compton upscattering and by additional bremsstrahlung emission.
The spectrum above 6~keV exhibits Compton-broadened K$\alpha$ emission and
K-absorption features due to Fe\,{\sc xxv}. The characteristic ``reflection hump''
produced by Compton downscattering of high-energy photons is apparent
at $E\sim 30\,{\rm keV}$.

\section{Constant-density models}

In reality, the detailed geometry and boundary conditions of the
illuminated disc surface are poorly known.  Therefore, it makes little
sense to pursue hydrostatic-equilibrium models like the one presented
in \S 2 any further. To study a range of reflection spectra that are
possible, we simplify the situation by considering reflection by a
constant-density atmosphere (see Ross \& Fabian 2005). Indeed,
Ballantyne et al.\ (2001) found that even under AGN conditions, models
for reflection by atmospheres in hydrostatic equilibrium could be
fitted by diluted (smaller reflection fraction) versions of
constant-density reflection models.

The atmosphere of the accretion disc is approximated by a slab with fixed hydrogen density 
$n_{\rm H}=10^{20}\,{\rm cm}^{-3}$ and Thomson depth $\taut=10$. The thermal emission 
from the disc is represented by a blackbody spectrum with designated temperature 
$T_{\rm BB}$ entering the slab from below. The outer surface is illuminated by a 
cutoff power-law spectrum with designated photon index $\Gamma$ and $F_0/F_{\rm disc}$
ratio. The $e$-folding energy for the high-energy cutoff to the illumination is fixed at
$E_{\rm cut} = 300\,{\rm keV}$, and the illumination again has a abrupt low-energy cutoff 
at $E = 0.1\,{\rm keV}$.

\begin{figure}
\includegraphics[scale=0.34,angle=270]{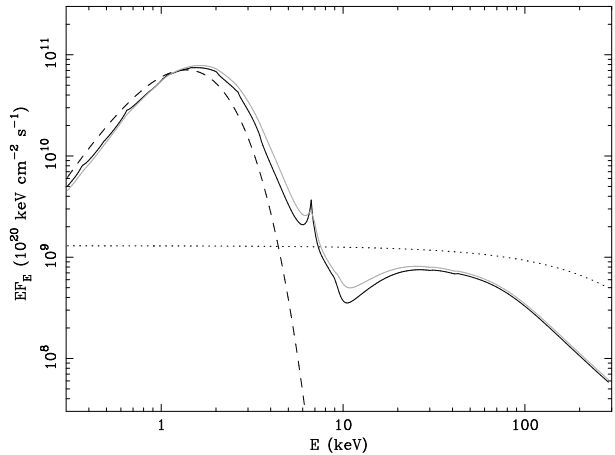}
\caption{Constant-density reflection model. The darker solid curve shows the emerging spectrum
when the thermal emission from the disc (dashed) has $kT_{\rm BB}=0.35\,{\rm keV}$, while
the illumination (dotted) has $\Gamma=2$ and $F_0=0.1F_{\rm disc}$. For comparison, the lighter
solid curve shows the spectrum from Fig.~2.} 
\end{figure}
Figure~3 shows the emerging spectrum for a constant-density model with the same parameters as 
the hydrostatic-atmosphere model discussed previously, namely $kT_{\rm BB}=0.35\,{\rm keV}$, 
$\Gamma=2$, and $F_0/F_{\rm disc}=0.1$. The constant density chosen corresponds to the
density for $\taut\ga 1$ in the hydrostatic-atmosphere model, and the temperature in the 
constant-density model agrees well for $\taut\ga 1$. Since the density is higher for $\taut\la 1$, 
the effective ionization parameter is lower, and the gas is somewhat cooler and less highly ionized 
in the constant-density model. As a result, the Fe K$\alpha$ line is slightly stronger, the iron K-edge 
is somewhat deeper, and at lower energies, tiny emission features are produced by recombinations to 
hydrogen-like ions of the lighter elements. Overall, however, the emergent spectrum is seen to be in 
reasonably good agreement with the result for an atmosphere in hydrostatic equilibrium.

\begin{figure}
\includegraphics[scale=0.34,angle=270]{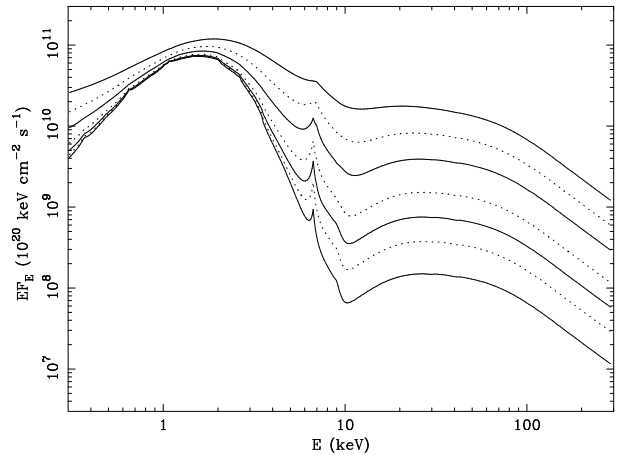}
\caption{Emergent spectra when $F_0/F_{\rm disc}=0.02$ (bottom solid curve), 0.05~(dotted), 
0.1~(solid), 0.2~(dotted), 0.5~(solid), 1~(dotted) and 2 (top solid curve). Here 
$kT_{\rm BB}=0.35\,{\rm keV}$ and $\Gamma=2$ throughout.}
\end{figure}
The effect of modifying the total illuminating flux is shown in Figure~4. Of course, the
thermal radiation becomes less dominant as the $F_0/F_{\rm disc}$ ratio increases.
For the four lowest values of $F_0/F_{\rm disc}$ shown in Fig.~4, the illumination is
so weak that the thermal radiation determines the ionization structure, with Fe\,{\sc xxv}
dominating throughout the atmosphere. For $F_0 = 0.5F_{\rm disc}$, however, the illumination
has an ionization parameter $\xi_0 = 970$ erg\,cm\,s$^{-1}$, where
\begin{equation}
\xi_0=\frac{4\pi F_0}{n_{\rm H}}.
\end{equation}
This makes Fe\,{\sc xxvi} the dominant species near the illuminated surface 
($\taut\la 0.4$), and Fe\,{\sc xxvi} contributes significantly to the broad iron K$\alpha$ 
line. For $F_0 = F_{\rm disc}$, iron is fully ionized for $\taut\la 0.4$, and Fe\,{\sc xxvi}
dominates for $0.4\la\taut\la 1.7$. Then Fe\,{\sc xxv} and Fe\,{\sc xxvi} together produce a
particularly broad K$\alpha$ blend. For $F_0 = 2F_{\rm disc}$ (top curve in Fig.~4), 
iron is fully ionized for $\taut\la 2$, and the iron K$\alpha$ line has weakened.

\begin{figure}
\includegraphics[scale=0.34,angle=270]{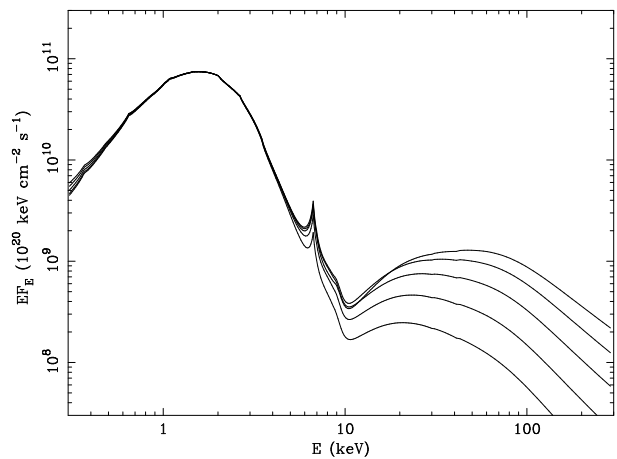}
\caption{Emergent spectra when $\Gamma = 1.6$ (top curve at high photon energies), 
1.8, 2.0, 2.2 and 2.4 (bottom curve at high photon energies). Here 
$kT_{\rm BB}=0.35\,{\rm keV}$ and $F_0/F_{\rm disc}=0.1$ throughout.}
\end{figure}
The effect of varying the power-law index $\Gamma$ is shown in Figure~5. The main
effect is on the high-energy tail above $\sim\!10\,{\rm keV}$, where Compton
downscattered illumination dominates the spectrum. For the steepest
illuminating spectrum ($\Gamma=2.4$), the 6--10 keV portion of the emergent
spectrum is also lowered.

\begin{figure}
\includegraphics[scale=0.34,angle=270]{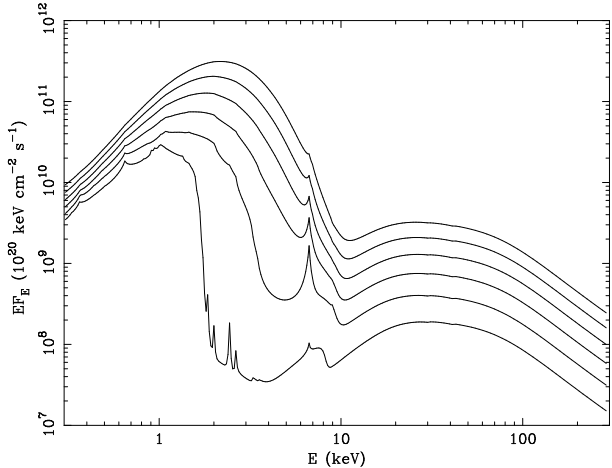}
\caption{Emergent spectra when $kT_{\rm BB} = 0.25$ (bottom curve), 
0.30, 0.35, 0.40, 0.45 and 0.50 keV (top curve). Here 
$\Gamma = 2$ and $F_0/F_{\rm disc}=0.1$ throughout.}
\end{figure}
Finally, Figure~6 shows the effect of varying the blackbody temperature 
of the disc emission. For $kT_{\rm BB} = 0.25\,{\rm keV}$, the thermal
emission is too soft and weak to keep heavier elements fully ionized.
Here Mg\,{\sc xii}, Si\,{\sc xiii}--{\sc xiv}, S\,{\sc xv}, and 
Fe\,{\sc xxi}--{\sc xxiii} dominate for $\taut\la 1$. There is a steep
dropoff in the emergent spectrum due to iron L-shell absorption, and
K$\alpha$ lines of Si\,{\sc xiii}--{\sc xiv} and S\,{\sc xv}--{\sc xvi}
are visible. The iron K$\alpha$ line is weak due to the destruction
of these photons via the Auger effect during resonant trapping
(see Ross, Fabian \& Brandt 1996). For $kT_{\rm BB} = 0.30\,{\rm keV}$,
S\,{\sc xvi} and Fe\,{\sc xxv} dominate near the illuminated surface,
and the emergent spectrum exhibits a strong Fe\,{\sc xxv} K$\alpha$ line.
As $kT_{\rm BB}$ increases beyond 0.35~keV, Fe\,{\sc xxv} remains the dominant 
species in the atmosphere, but the iron K$\alpha$ line becomes weaker and 
more difficult to distinguish from the strengthening thermal continuum.

\section{Discussion}

In the high/soft state of a BHB, the X-ray spectrum is dominated by
the soft thermal radiation, which generally accounts for $\ga\!75\%$
of the flux in the 2--20~keV range. Additional spectral components are
generally a high-energy tail, which is approximately a power law with
photon index $\Gamma\sim 2$, and a broad iron K$\alpha$ emission
feature (see Tanaka \& Lewin 1995; Reynolds \& Nowak 2003; McClintock
\& Remillard 2006; Remillard \& McClintock 2006). We have demonstrated
how illumination of a BHB accretion disc can help produce such features.

\begin{figure}
\includegraphics[scale=0.34,angle=270]{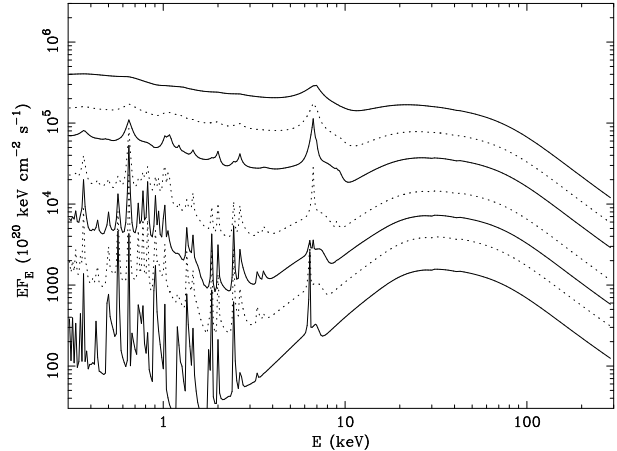}
\caption{Constant-density reflection spectra under AGN conditions: the
atmosphere has $n_{\rm H}=10^{15}\,{\rm cm}^{-3}$, and the disc below
is cold. Emergent spectra are shown for $\Gamma=2$ illumination with the 
same ionization parameters as in Fig.~4, namely $\xi_0=38.8$ (bottom solid 
curve), 97.0~(dotted), 194~(solid), 388~(dotted), 970~(solid), 
1940~(dotted), and 3880 erg\,cm\,s$^{-1}$ (top solid curve).}
\end{figure}
The emergent spectra under black-hole binary conditions are quite 
different than under the conditions in active galactic nuclei (AGN).
For comparison, Fig.~7 shows the corresponding reflection spectra
under conditions that would be expected in an active galactic nucleus. 
The density in the accretion-disc atmosphere and the total
illuminating flux are lower, and thermal radiation from the cooler 
disc, which lies well below X-ray energies, is ignored. The models
calculated have the same values for the illumination ionization parameter,
$\xi_0$, as those shown in Fig.~4. With lower temperatures and little 
ionizing radiation within the disc, the metals are not highly ionized. 
For $\xi=38.8$ and 97.0 erg\,cm\,s$^{-1}$, the atmosphere is only highly 
ionized in a very thin surface layer. Now the iron K$\alpha$ emission is 
dominated by the 6.4-keV fluorescence of weakly ionized atoms. Strong 
emission features at lower energies are due to K-lines of the lighter
elements as well as L-lines of iron, while the continuum there is
suppressed by bound-free absorption. For $\xi=194$ erg\,cm\,s$^{-1}$, 
the iron K$\alpha$ line is weak due to resonant Auger destruction. For
$\xi_0=388$ and 970 erg\,cm\,s$^{-1}$, the Fe~{\sc xxv} K$\alpha$ line 
at 6.7~keV dominates, but strong emission lines are still present at
lower energies. For $\xi=1940$ and 3880 erg\,cm\,s$^{-1}$, the gas
is highly ionized within a deep surface layer, and the reflected
spectrum is fairly similar to the BHB case, except that the lack of 
soft thermal radiation makes the spectrum flatter for $E\la 10\,{\rm keV}$.

\begin{figure}
\includegraphics[scale=0.34,angle=270]{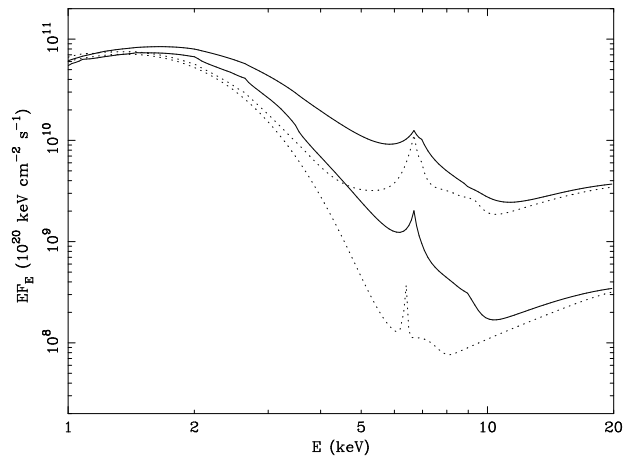}
\caption{Solid curves show the emergent spectra for BHB models with 
$kT_{\rm BB}=0.35\,{\rm keV}$, $\Gamma=2$, and $F_0/F_{\rm disc}$ values
of 0.05 (lower) and 0.50 (upper). Dotted curves show the result of adding a 
0.35-keV blackbody spectrum to AGN reflection spectra with $\xi_0$ values of
97.0 (lower) and 970\,erg\,cm\,s$^{-1}$ (upper) that have been increased by a 
factor of $10^5$.}
\end{figure}
In particular, reflection in a BHB system cannot be mimicked simply by
adding a blackbody spectrum to the reflection spectrum from an otherwise
cool disc. This is demonstrated in Figure 8, which shows two examples of 
adjusting a local AGN reflection spectrum upward by a factor of $10^5$ and then
adding a blackbody spectrum to represent the emission by the disc. For
$\xi_0=97.0$ erg\,cm\,s$^{-1}$, the mimicked spectrum has a narrow iron line
at 6.4~keV, while the actual BHB spectrum has a strong, Compton-broadened 
Fe\,{\sc xxv} line (6.7 keV) atop a much higher continuum. For 
$\xi_0=970$ erg\,cm\,s$^{-1}$, on the other hand, the mimicked spectrum does have
a broad Fe\,{\sc xxv} line, but the actual BHB spectrum has a very different line
shape due to a blend of Fe\,{\sc xxv} and Fe\,{\sc xxvi} emission, and the
surrounding continuum is again higher.

If the accreting object was a neutron star ($M\approx 1.4M_{\odot}$) instead
of a 10-$M_{\odot}$ black hole, then both $F_{\rm disc}$ and $n_{\rm H}$
would be $\sim\!7$ times larger for the same values of $\varepsilon$ and
$R/R_{\rm g}$. Although the ``ionization parameter'' of the disc radiation 
would be the same, its harder spectrum ($kT_{\rm BB}$ about 1.6 times larger) 
would give it greater ionizing power.  

\begin{figure}
\includegraphics[scale=0.34,angle=270]{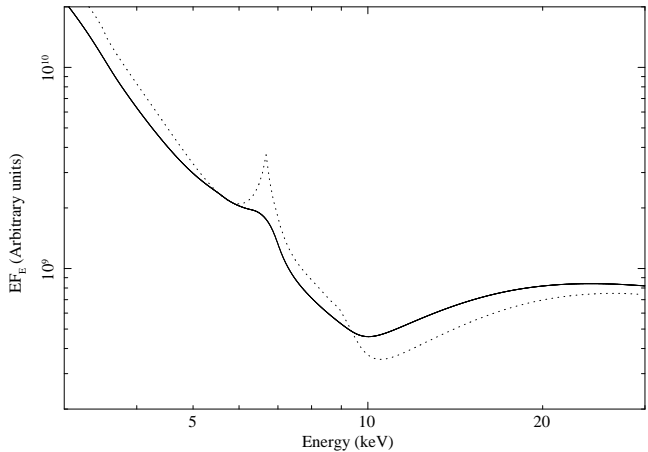}
\caption{Effect of relativistic blurring on the reflected spectrum. A model
with $kT_{\rm BB}=0.35\,{\rm keV}$, $\Gamma=2$, and $F_0/F_{\rm disc}=0.1$ 
(dotted curve) has been blurred (solid curve) as if observed from a region 
with emissivity index 3 extending from 6 to 20 gravitational radii in a disc 
around a rotating black hole, and viewed at an inclination angle of $30^{\circ}$.} 
\end{figure}
Of course, spectral features due to reflection by the inner portion of
the accretion disc will be blurred by the relativistic Doppler effect
and gravitational redshift. Figure~9 shows an example of such
relativistic blurring for the model with $kT_{\rm BB}=0.35\,{\rm keV}$,
$\Gamma=2$, and $F_0/F_{\rm disc}=0.1$ discussed in \S 3.  The Fe~{\sc xxv}
K$\alpha$ line, already broadened by Compton scattering, is further 
broadened and redshifted. The iron K-edge in the reflected spectrum is
also smeared out, making it less prominent (see Ross et al.\ 1996). 
In this way it mimics the {\tt smedge} model in XSPEC often
used for fitting the spectra of BHB. This model is a simple
phenomenological representation of a smeared absorption edge (Ebisawa
et al.\ 1994), and has no clear physical basis. 

Compton broadening of the iron K$\alpha$ line is relatively more important 
for BHB at a given $\xi_0$ value than for AGN since the surface layers are 
hotter. This is important when deducing the spin of the black hole from the 
overall observed breadth of the line.

\section{Acknowledgments} 
We thank Jon Miller and an anonymous referee for helpful comments.
RRR and ACF thank the College of the Holy Cross and the Royal Society,
respectively, for support.


\bsp 

\label{lastpage}

\end{document}